\documentclass[aps,twocolumn]{revtex4}

\usepackage{graphicx}
\usepackage{amssymb, amsmath,mathtools}
\usepackage{bbm}

\graphicspath{{./}{./plots/}}

\DeclareMathOperator{\Tr}{Tr}
\DeclareMathOperator{\sign}{sign}

\DeclareMathOperator{\artanh}{artanh}

\begin{document}

\title{Nematic quantum criticality in three-dimensional Fermi system with
quadratic band touching}

\author{Lukas Janssen and Igor F. Herbut}

\affiliation{Department of Physics, Simon Fraser University, Burnaby, British
Columbia, Canada V5A 1S6}

\begin{abstract}

We construct and discuss the field theory for tensorial nematic order parameter
coupled to gapless four-component fermions at the  quadratic band touching point
in three (spatial) dimensions. Within a properly formulated epsilon-expansion
this theory is found to have a quantum critical point, which describes the
(presumably continuous) transition from the semimetal into a (nematic) Mott
insulator. The latter phase breaks the rotational, but not the time-reversal
symmetry, and  may be relevant to materials such as gray tin or mercury
telluride at low temperatures. The critical point represents a simple
quantum analogue of the familiar classical isotropic-to-nematic transition in
liquid crystals. The properties and the consequences of this quantum critical
point are discussed. Its existence supports the scenario of the ``fixed-point collision'',
according to which three-dimensional Fermi systems with quadratic band touching
and long-range Coulomb interactions are unstable towards the gapped nematic
ground state at low temperatures.

\end{abstract}

\maketitle

\section{Introduction}

Electronic systems that have their Fermi surface reduced to Fermi points have
received plenty of attention lately. In particular, recent progress on the
problem of interacting Dirac electrons, when the dispersion near the Fermi
points is linear in momentum, has indicated that these systems suffer a quantum
phase transition with increasing interactions into a gapped phase, described
well by the relativistic field theory of the Gross-Neveu-Yukawa
type~\cite{parisen}. A weak long-range component of the Coulomb interaction
appears to be an irrelevant perturbation at the quantum critical (QC) point, and
the transition is essentially due to some of its short-range components becoming
sufficiently large. When the dispersion near the Fermi point(s) is quadratic, on
the other hand, the result is rather  different. In the bilayer graphene, for
example, it is one of many possible mass-gaps that opens up already at an
infinitesimal interaction~\cite{vafek}. The finite density of states that
accompanies such a quadratic band touching (QBT) in two dimensions (2D) causes
the long-range Coulomb interaction, loosely speaking, to be screened, and at the
same time the non-interacting ground state to be unstable at weak short-range
interaction~\cite{sun, dora}.

The situation in three-dimensional (3D) systems with QBT is maybe more
interesting. QBT arises naturally in many gapless semiconductors, such as gray
tin, mercury telluride, or certain pyrochlore iridates~\cite{krempa}, that
feature band inversion due to the spin-orbit coupling. The density of states at
the QBT point now vanishes, and the long-range nature of the electron-electron
interaction must be taken into account. It has been argued by Abrikosov long ago
\cite{abrikosov}, that the plain vanilla density-density Coulomb interaction in
a 3D system with the QBT should turn the ground state into an example of a
scale-invariant non-Fermi liquid (NFL). Such an exotic zero temperature phase
would manifest itself in characteristic nontrivial power laws in temperature or
frequency in various response functions of the system~\cite{moon}.

We have recently pointed out~\cite{janssen}, on the other hand, that a 3D system
with the chemical potential at the QBT and the Coulomb repulsion between the
electrons may be unstable towards an insulating ground state with an anisotropic
gap in the spectrum at low temperatures.
The mechanism responsible for this instability was proposed to be the collision
between the Abrikosov's infrared stable NFL fixed point with another, QC point,
which approaches it from the strong-coupling region as the spatial
dimensionality of the system is taken to be decreasing from $d=4$. The collision
of fixed points has been studied as a mechanism behind several interesting
instabilities in a variety of many-body systems in the past~\cite{zlatko1,
zlatko2, kaveh, gies, kaplan}. Within the standard one-loop calculation it
occurs here somewhat above and close to $d=3$, when both the NFL and the QC
fixed points become complex and disappear from the physical space of real
couplings. As a result, the coupling constants in the theory run away towards
the values at which spontaneous breaking of the rotational symmetry appears to
be the most favorable instability. The system in its interacting ground state
would effectively appear as if it were under, in this case {\it dynamically
generated}, strain.  Furthermore, in the materials with the rest of the band
structure equivalent to that of gray tin or mercury telluride as well, the
resulting insulating ground state, at least at the mean-field level, would be
topologically nontrivial~\cite{fu}. It would therefore be a precious example of
a {\it topological} Mott insulator~\cite{raghu, daghofer, grushin, duric}.

In order to remove the Abrikosov's NFL fixed point, however, the existence of
which is guaranteed close to four spatial dimensions, from the physical
real-valued space of couplings, it is necessary to have a QC point that would
collide with it with the change of some parameter. Indeed, in a certain
large-$N$ extension of the theory one can show that such a QC point does
exist~\cite{janssen}. At the physical value of $N=1$, however, in the purely
fermionic formulation of the problem the putative QC point lies at strong values
of the short-range couplings in the relevant dimensions $3 \leq d<4$. One may
therefore question whether such a QC point is a genuine feature of the theory,
and if it would, for example,  survive if one went beyond the one-loop
approximation. As we will see such reservations would not be entirely without
grounds. Similar issue arises in the interacting system of linearly dispersing
Dirac fermions~\cite{herbut1, herjurroy}. In this case, however, an alternative
partially bosonized Gross-Neveu-Yukawa formulation can be
devised~\cite{herjurvaf}. In this reformulation of the theory one finds a
clearly identifiable upper critical dimension, which can be used to control the
quantum critical point and compute its characteristics in perturbative fashion.
The crucial ingredient, however, behind this fortunate outcome is the linearity
of the Dirac quasiparticle spectrum, which allows the Lorentz symmetry, although
absent at the level of the lattice Hamiltonian, to {\it emerge} dynamically at
the QC point. In the systems with the QBT, on the other hand, such an enlarged
symmetry is certainly not expected at low energies, and it is {\it a priori} not
even clear
what dynamic scaling to assume, as the coupled fermion-boson system on the
Gaussian level appears to be characterized by two different dynamical critical
exponents, $z = 1$ and $2$, respectively.

Furthermore, as already implicit in~\cite{abrikosov} and as will be discussed
here at length, one readily finds that the minimal Hamiltonian with the QBT
point in 3D requires the use of the maximal set of {\it five} four-dimensional
mutually anticommuting Dirac matrices. This is not an accident, and the
situation is the same in 4D, except that one there needs the maximal set of nine
sixteen-dimensional Dirac matrices. Having no further anticommuting matrix left
prohibits then the opening of an isotropic mass-gap in the insulating state,
which is usually preferred in the systems with Dirac fermions~\cite{chamon,
isospin}. This leaves as the energetically next-best option the dynamical
generation of the second-rank tensorial order parameter, which breaks the
rotational and  preserves the time-reversal symmetry. Such a {\it nematic} order
parameter, as well known from the studies of liquid crystals~\cite{chaikin},
allows a cubic rotationally-invariant term, which is typically responsible for a
discontinuous transition. This makes the existence of the QC point in this
system seem additionally questionable.

Given these difficulties which appear to be inherent to the problem at hand, it
is quite remarkable that together they conspire to allow the construction of the
Gross-Neveu-Yukawa type of field theory for the nematic transition in the system
with QBT that {\it has} a perturbatively accessible QC point. We find that it is
precisely the presence of the cubic invariant for the nematic order parameter
that implies the existence of the upper critical dimension in the theory.
The dichotomy in the dynamical scaling of fermions and bosons at the Gaussian fixed point of the theory
is naturally resolved at the QC point, with the critical behavior becoming independent of the
specific choice of the scaling scheme, and ultimately characterized by a single
dynamical critical exponent $z$.
The nematic quantum phase transition from the semimetallic phase into the insulating
phase with the anisotropic gap described by the above QC point is therefore presumably 
\emph{continuous}, at least at the level of the mean-field theory---in contrast to
the classical thermal isotropic-to-nematic transition in liquid
crystals~\cite{chaikin}.
 To the leading order, the QC point is characterized by
the dynamical critical exponent $z=2$, a nontrivial positive anomalous dimension
of the order-parameter field, and a vanishing anomalous dimension for the
fermions. The relative signs of the cubic-term and the Yukawa couplings
at the critical point are also such that the state with fully gapped fermions is
favored in the ordered phase, as one would expect from
energetics~\cite{janssen}.

For the sake of simplicity, in the present work we neglect the effects of
the unscreened long-range tail of the Coulomb interaction. While our predictions
for the critical exponents near the upper critical dimension may be subject to
quantitative improvement upon its inclusion, possibly already at the leading order
in the $\epsilon$~expansion, we see no reason why the mechanisms responsible
for the \emph{existence} of the nematic QC point should be qualitatively
altered when the long-range Coulomb interaction is included---as long as
$\epsilon$ is small. For larger values of $\epsilon$, in contrast, according to
the scenario of the ``fixed-point collision''~\cite{janssen}, the QBT system
with the
long-range interaction should become unstable towards an insulating ground
state with an anisotropic gap. With the inclusion of the long-range interactions the expansion around the upper
critical dimension is thus expected to, even qualitatively, eventually break down at some
\emph{lower} critical dimension, which may as well lie above the physical
three~\cite{janssen}. Rather than deriving quantitative predictions for
experimental systems, our limited objective here is thus to further
substantiate the mechanism  of ``fixed-point collision", by establishing the
very existence of the nematic critical point beyond the previous large-$N$
approximation.

The organization of the paper is as follows. In the next section we discuss the
construction of the minimal isotropic QBT Hamiltonian in the form that most
closely resembles the Dirac Hamiltonian, in general spatial dimension.  In
Sec.~\ref{sec:GNY-field-theory} the Gross-Neveu-Yukawa continuum field theory
for the nematic order parameter coupled to fermions is presented. We present the
mean-field theory for the nematic quantum phase transition and discuss its order
and the nature of the associated interacting ground state in
Sec.~\ref{sec:mf-theory}.
The structure of the renormalization group and the concomitant quantum critical
point are discussed in Sec.~\ref{sec:rg-flow}. In Sec.~\ref{sec:interpretation}
we offer an interpretation of  our results. Concluding remarks are given in
Sec.~\ref{sec:conclusions}. Some nontrivial technical points necessary for the
calculation are presented in five appendices.

\section{QBT Hamiltonians in different dimensions}

We first discuss the construction of the minimal, rotationally invariant and
particle-hole symmetric QBT Hamiltonian, in general spatial dimension $d$. We
assume that in the momentum representation it has the form of
\begin{equation}
H= \sum_{i,j=1}^d G_{ij} p_i p_j,
\end{equation}
with $G_{ij}$ as the matrix coefficients, which need to be determined.
Obviously, $G_{ij}$ must transform as the components of a second-rank symmetric
tensor under rotations. For simplicity, we set the effective band mass to
$2m=1$ and demand that $H^2 = p^4 \mathbbm 1$, with the minimal dimension of
the Hamiltonian to be determined. $H^2$ thus contains only even powers of the
momentum's components $p_i$, and the matrix coefficients must then satisfy the
anticommutation rules
\begin{equation}
\{ G_\mathrm{of}, G_\mathrm{of}' \} = \{G_\mathrm d, G_\mathrm{of} \}= 0,
\end{equation}
where $G_\mathrm d$ is any of the diagonal elements $G_{ii}$, $G_\mathrm{of}$ is
any of the off-diagonal element $G_{ij}$ with $i\neq j$, and $G_\mathrm{of} \neq
G_\mathrm{of}'$. Then
\begin{equation}
H^2 = \sum_{i=1}^d G_{ii}^2 p_i ^4 + \sum_{i<j} p_i ^2 p_j ^2 \left(4 G_{ij} ^2 + \{G_{ii}, G_{jj} \} \right).
\end{equation}
If we normalize the diagonal elements so that all $G_\mathrm d^2 = 1$,
$H^2 = p^4$ provided that the following condition is satisfied:
\begin{equation} \label{eq:condition-Gij-A}
4 G_\mathrm{of}^2 + \{G_\mathrm d, G_\mathrm d '\} =2.
\end{equation}
Demanding further that the tensor $G_{ij}$ is traceless, the Hamiltonian $H$
would contain only the irreducible tensor $p_i p_j - \delta_{ij}p^2 /d$, and
would be without the scalar term $\propto p^2$. The existence of such a scalar
part would only introduce different curvatures of the upper and the lower
branches of the energy spectrum, and we omit it for the time being. We therefore
set
\begin{equation}
\sum_{i=1}^d G_{ii} =0,
\end{equation}
with the spectrum being quadratic, isotropic, and particle-hole symmetric,
$\varepsilon_\pm(\vec p) = \pm p^2$.
Tracelessness, however, implies that, for arbitrary index $k$,
\begin{equation}
0 = \{ G_{kk}, \sum_{i=1}^d G_{ii} \} = 2+ \sum_{i (\neq k) } \{ G_{kk},G_{ii}\}
\end{equation}
or, in other words that for any pair of diagonal elements
\begin{equation} \label{eq:condition-Gij-B}
\{ G_\mathrm d, G_\mathrm d' \} = \frac{2}{1-d}.
\end{equation}
When combined with Eq.~\eqref{eq:condition-Gij-A} this in particular implies
that off-diagonal elements are to be normalized as $G_\mathrm{of}^2 =
d/(2(d-1))$.

To construct the desired Hamiltonian $H$ we therefore need
\begin{equation}
\left( \frac{d^2 - d}{2} \right) + (d-1)
\end{equation}
mutually anticommuting Dirac matrices, for the off-diagonal (first) and the
diagonal (second term) elements. Out of $d-1$ Dirac matrices for the diagonal
matrices, $d$ matrices $G_{ii}$ that satisfy Eq.~\eqref{eq:condition-Gij-B} and
square to unity can always be constructed.

For example:

(1) In $d=2$ only two anticommuting matrices are needed, and therefore may be
chosen as $G_{12}=G_{21}= \sigma_1$, and $G_{11}=-G_{22}= \sigma_3$. The
Hamiltonian describes the band touching point in bilayer graphene, for example.
Note that $H$ is time-reversal symmetric, and the time-reversal operator is
$T=K$, the complex conjugation alone. Since $T^2 =1$ this Hamiltonian can arise
as a low-energy limit of a lattice Hamiltonian with spinless fermions hopping
between sites~\cite{herbut2}.  Examples of such lattice Hamiltonians already
exist in the literature~\cite{sun, dora}.

(2) In $d=3$ one needs {\it five} Dirac matrices for the construction, so their
minimal dimension is four. We can choose $G_{12}= (\sqrt{3}/2) \gamma_2$,
$G_{13}= (\sqrt{3}/2) \gamma_3$, $G_{23}= (\sqrt{3}/2) \gamma_4$, and then for
the diagonal elements
\begin{align}
G_{11} & =  - \frac{1}{2} \gamma_5 + \frac{\sqrt{3}}{2} \gamma_1, \\
G_{22} & =  - \frac{1}{2} \gamma_5 - \frac{\sqrt{3}}{2} \gamma_1, \\
G_{33} & =\gamma_5.
\end{align}
The Hermitian Dirac matrices $\gamma_a$, $a=1,\dots, 5$ satisfy the Clifford
algebra $\{\gamma_a, \gamma_b \} = 2\delta_{ab}$. With this particular choice
the Hamiltonian can also be rewritten as
\begin{equation} \label{eq:hamiltonian-spherical-harmonics}
H= \sum_{a=1}^5 d_a (\vec{p} ) \gamma_a,
\end{equation}
with $ d_a (\vec{p})=p^2 \tilde{d}_a (\theta,\varphi)$ are proportional to five
real spherical harmonics for the angular momentum of {\it two}; explicitly,
$\tilde{d}_1 + i \tilde{d}_2 = (\sqrt{3}/2)\sin^2 (\theta) e^{2i\varphi}$,
$\tilde{d}_3 + i \tilde{d}_4 = (\sqrt{3}/2)\sin (2\theta) e^{i\varphi}$,
$\tilde{d}_5= (3 \cos^2 \theta -1)/2$, with $\theta$ and $\varphi$ as the
spherical angles in the momentum space.

Note that among the five four-dimensional Dirac matrices we can always choose
two (say $\gamma_4$ and $\gamma_5$) as imaginary and the remaining three as
real, so $H$ is also time-reversal invariant, but now with (unique) $T=\gamma_4
\gamma_5 K$~\cite{herbut2}. Most importantly, $T^2=-1$, and in three dimensions
$H$ inevitably  describes particles with half-integer spin. In fact this
``Luttinger Hamiltonian" is well known to arise from the spin-orbit coupling in
gapless semiconductors such as gray tin, for example~\cite{luttinger, murakami},
Also, the Kramers' theorem applies in this case and dictates that the spectrum
is doubly degenerate at any  momentum.

(3) For completeness, let us also display the  solution for $d=4$. For the
off-diagonal elements we now need six mutually anticommuting matrices, and for
the diagonal elements three more. The nine-component Clifford algebra has the
unique irreducible  representation being sixteen-dimensional. We may then choose
the off-diagonal elements as $(G_{12}, G_{13}, G_{23}, G_{14}, G_{24}, G_{45}) =
\sqrt{2}/3 (\gamma_2, \gamma_3, \gamma_4, \gamma_6, \gamma_7, \gamma_8)$, and
the diagonal elements as
\begin{align}
G_{11} & = -\frac{1}{3} \gamma_9 - \frac{\sqrt{2}}{3} \gamma_5 +  \sqrt{ \frac{2}{3}} \gamma_1, \\
G_{22} & = -\frac{1}{3} \gamma_9 - \frac{\sqrt{2}}{3} \gamma_5 - \sqrt{ \frac{2}{3}} \gamma_1, \\
G_{33} & = -\frac{1}{3} \gamma_9 + \frac{\sqrt{8}}{3} \gamma_5, \\
G_{44} & = \gamma_9.
\end{align}
Displaying $H$ in the form equivalent to
Eq.~\eqref{eq:hamiltonian-spherical-harmonics} would define the four-dimensional
generalization of the $\ell=2$ spherical harmonics. Note also that among the
nine sixteen-dimensional Dirac matrices, four (say with indices $a=6,7,8,9$) can
be chosen to be purely imaginary, with the remaining five then as
real~\cite{herbut3}. The time-reversal operator that commutes with $H$ exists,
and is unique: $T=\gamma_6 \gamma_7 \gamma_8 \gamma_9 K$, but again $T^2=+1$,
and the minimal Hamiltonian in $d=4$, similarly to $d=2$ describes a spinless
particle.

The solutions to the above conditions for the matrices $G_{ij}$ can be found in
all dimensions, with the properties of the minimal Hamiltonian under time
reversal, for example, being strongly dimension dependent, as our examples
already illustrate. Further details on the construction of the $d$-dimensional
QBT Hamiltonian are provided in Appendix~\ref{app:QBT-general}. The construction
can also be generalized to higher-order band touching, which would involve the
higher-rank tensors and higher-angular-momentum spherical harmonics. Further
elaboration of this point would be somewhat tangential to our main subject, and
we leave it for another occasion.

\section{The Gross-Neveu-Yukawa field theory}
\label{sec:GNY-field-theory}

We consider next the QBT fermions in $d=3$ and at $T=0$.
The system may possibly harbor several QC fixed points which for large enough
short-range interactions could lead to various different instabilities and
corresponding symmetry breaking patterns, in analogy to the 2D Dirac system
describing interacting fermions on the honeycomb lattice~\cite{herjurroy}. Due
to the vanishing density of states at the QBT point in 3D any such QC point will
be located at strong coupling---as long as the long-range tail of Coulomb
interaction is suppressed. With the long-range interaction included, however,
the critical coupling is expected to decrease significantly, and might even
vanish completely~\cite{janssen}, in
contrast to the Dirac systems. We have shown recently~\cite{janssen}, that in
the isotropic and particle-hole symmetric case the long-range interaction favors
the nematic instability, and if indeed the QBT point becomes unstable at low
temperatures, the rotational symmetry would break
spontaneously. To substantiate this scenario and to understand to concomitant
ordering, in the present work we therefore focus on the nematic interaction
channel.

In order to establish the existence of a nematic QC point and to discuss
its characteristics, we will suppress in what follows
the long-range part of the Coulomb interaction. In the vicinity of the upper
critical dimension, its inclusion is expected to only \emph{quantitatively}
improve our numerical predictions. Away from the upper critical dimension,
however, the situation may, according to the scenario of the ``fixed-point
collision''~\cite{janssen}, dramatically change, and we will briefly comment on
this in the conclusions.
The continuum quantum action, coupled to the fluctuating
nematic order parameter then is $S= \int d\tau d^d \vec{x} L$, with the
Lagrangian density
\begin{equation} \label{eq:L}
L =  L_\psi + L_{\psi \phi} + L_\phi,
\end{equation}
and with the individual terms defined as
\begin{align}
L_\psi & = \psi^\dagger \left( \partial_\tau + \gamma_a d_a (-i \nabla)  \right) \psi, \\
L_{\psi \phi} & = g \phi_a \psi^\dagger \gamma_a \psi, \\
L_\phi & = \frac{1}{4} T_{ij} \left(- c \partial_\tau^2 - \nabla^2 + r \right)
T_{ji} + \lambda T_{ij} T_{jk} T_{ki} \nonumber \\
&\quad + \mathcal O(T^4). \label{eq:Lphi}
\end{align}
$\psi$ is the four-component Grassmann field, whereas $\phi_a$ is a real field.
The summation over the repeated indices is now assumed, and  $a=1,\dots, 5$, and
$i,j,k=1,2,3$. $\gamma_a$ are the five mutually anticommuting four-dimensional
Dirac matrices introduced earlier.

The real, symmetric, traceless tensor field $T_{ij}$ is defined as
\begin{equation}
T_{ij}= \phi_a \Lambda_{a,ij},
\end{equation}
where $\Lambda_a$ are the five real, symmetric, three-dimensional Gell-Mann
matrices.
Their explicit form and important properties are discussed in
Appendix~\ref{app:Gell-Mann-matrices}.  Since the five spherical harmonics
$d_a(\vec{p})$ transform as the components of the traceless symmetric tensor of
rank two under rotations, the Lagrangian $L$ will be invariant under rotations
provided that the five components of the tensor $T_{ij}$, $\phi_a$,
$a=1,\dots,5$ do so as well. At the level of the quantum mechanical averages
\begin{equation}
\langle \phi_a \rangle = \frac{-g}{r} \langle \psi^\dagger \gamma_a \psi \rangle,
\end{equation}
and finding $\langle \phi_a \rangle\neq 0 $ signals spontaneous breaking of the
rotational symmetry.  The tensor $T_{ij}$ can be understood as the {\it nematic}
 order parameter, in analogy with liquid crystals, where the identical object
describes the finite temperature phase transition between the isotropic and
anisotropic phases~\cite{chaikin}.

In the context of 2D metals, a nematic QC point describing the $\ell=2$
Pomeranchuk instability of Fermi-liquid theory has been thoroughly investigated
previously~\cite{oganesyan, metlitski} and still commands
attention~\cite{holder}, also due to its potential role in the phase diagram of
certain high-temperature superconductors~\cite{fradkin}. Nematic instabilities
have also been predicted in 2D Fermi systems with QBT~\cite{vafek, sun, dora}.
In two dimensions, however, the order parameter is odd under $\pi/2$ spatial
rotation, which forbids a cubic term $\propto \Tr T^3$ in the
action~\cite{oganesyan}. By contrast, the 3D system defined by
Eqs.~\eqref{eq:L}--\eqref{eq:Lphi} is an immediate generalization of the field
theory describing the classical nematic transition in liquid
crystals~\cite{chaikin}, which is recovered when the fermions are decoupled,
i.e., in the limit $g \to 0$. The critical point we will find shortly
therefore represents possibly the simplest quantum analogue of this familiar
classical nematic transition.

The above form of the Lagrangian $L$ contains the minimal number of parameters,
and the imaginary time, length, and the Grassmann and the real fields have been
rescaled so that the coefficients in front of the first and the second terms in
$L_\psi$, and the second term in $L_\phi$ are brought to unity. Besides the
tuning parameter $r$, this still leaves the coefficient in the first term in
$L_{\phi}$, $c$, and the two interaction coupling constants: Yukawa coupling
$g$,  and the cubic term self-interaction $\lambda$. These have the engineering
dimensions
\begin{equation} \label{eq:dim-g-lambda}
\dim[g]= \dim[\lambda]= \frac{6-z-d}{2},
\end{equation}
whereas
\begin{equation}
\dim[c]= 2-2z.
\end{equation}
Keeping the coefficients in $L_\psi$ fixed demands the dynamical critical
exponent to be $z=2$ at the Gaussian fixed point $\lambda=g=0$. One then finds
that both couplings $g$ and $\lambda$ become relevant in the infrared  {\it
simultaneously} below $d=4$. This observation allows one to formulate a
perturbative approach to the problem of the infrared behavior as the expansion
in the small parameter
\begin{equation}
\epsilon=4-d,
\end{equation}
and search for possible non-Gaussian critical points in the theory. The terms
$\mathcal O(T^4)$ in $L_\phi$ have for this reason been omitted as irrelevant to
the leading order in~$\epsilon$.
The parameter $c$ also is irrelevant at the Gaussian fixed point when
$z=2$. At the QC fixed point, however, we will find $c$ to be shifted to a finite
positive value, leading to nontrivial dynamical scaling of the order parameter
(see Sec.~\ref{sec:rg-flow}).

The rescaling procedure involves an apparent ambiguity, as one might
equally well fix the coefficient $c$ in front of the first term in $L_\phi$,
$\propto c (T_{ij} \partial_\tau^2 T_{ji})/4$, to unity, and let the
coefficient, let us call it $a$, in front of the first term in $L_\psi$,
$\propto a (\psi^\dagger \partial_\tau \psi)$,  to flow instead. This would
dictate a different dynamical exponent, $z=1$, at the Gaussian fixed point,
reflecting the fact that the noninteracting system possesses two different
characteristic time scales. In Appendix~\ref{app:dyn-scaling} we show that this
alternative prescription leads to an equivalent RG flow, and the
\emph{same} universal quantities at criticality. The QC point will thus be
characterized by a single diverging time scale and a unique dynamical exponent.
A similar ambiguity occurs in the effective order-parameter theory for
the nematic instability in 2D metals, when the fermions have been integrated
out, though its resolution differs from the present case~\cite{meng}.

We should also comment on yet another rotational invariant quadratic in
$T_{ij}$, which is proportional to
\begin{equation} \label{eq:beta-term}
\partial_i T_{ij} \partial_k T_{kj},
\end{equation}
and that we have omitted in $L_\phi$. It couples spatial rotations to internal
rotations of the nematic order parameter, and is thus possible only when the
dimension $p$ of the tensor $T_{ij}$ ($i,j=1,\dots,p$) is equal to the spatial
dimension $d$ (as is the case in our problem). We find that although of the same
engineering dimension as the term we included at the noninteracting fixed point,
this term develops a {\it negative} anomalous dimension to the leading order in
interactions, and as such we expect it to become irrelevant at the interacting
critical point (see Appendix~\ref{app:RG-flow-diagrams}). One can analogously
justify the common omission of this term in the studies of the classical
isotropic-to-nematic transition in three dimensions.

\section{Mean-field theory}
\label{sec:mf-theory}

Before we present the solution of the problem in the vicinity of the upper
critical dimension, let us consider the mean-field theory in which the
fluctuations of the order-parameter field $\phi_a$ are neglected. This
approximation can be justified by adding an additional ``flavor'' index to the
fermions (e.g., by allowing more than one QBT point at the Fermi level) and
taking the limit of large flavor number $N$~\cite{janssen}.
The mean-field theory is solved by minimizing the total energy
\begin{equation} \label{eq:mf-energy}
 E_\text{MF}(\phi_1,\dots,\phi_5) = \frac{r}{2} \phi_a \phi_a + 2 \int_0^\Lambda \frac{d \vec p}{(2\pi)^3} \varepsilon(\vec p)
\end{equation}
where $\varepsilon(\vec p)$ denote the lower-branch energy eigenvalues of
the mean-field Hamiltonian $H_\text{MF}(\vec p) = p^2 \tilde d_a(\theta,\varphi)
\gamma_a + g \phi_a \gamma_a$ in the presence of constant $\phi_a$, viz.,
\begin{equation}
 \varepsilon(p,\theta,\varphi) = - p^2 \sqrt{1 + 2 \tilde d_a(\theta,\varphi) \frac{g \phi_a}{p^2} + \left(\frac{g \phi_a}{p^2}\right)^2}.
\end{equation}
$\Lambda$ is the UV momentum cutoff, $0\leq |\vec p| \leq \Lambda$. For
convenience, and  without loss of generality, let us assume $g>0$.
The first term in Eq.~\eqref{eq:mf-energy} represents the energy cost of a
finite $\phi_a$. By contrast, the second term decreases with increasing order
parameter, and thus involves the energy gain due to a (possible) ordering. It
can be interpreted as the sum of the energies of the filled, doubly degenerate
single-particle states in the ordered phase, with the Fermi level at the QBT. In
the present model without the long-range Coulomb interaction and in $d=3$ we
expect the ordered state to be energetically favorable if the parameter $g^2/r$
exceeds a certain strong-coupling threshold. This threshold, however, may
decrease  substantially upon the inclusion of the long-range part of the Coulomb
repulsion, and might even vanish completely~\cite{janssen}.

In the reference frame in which the tensor order parameter becomes diagonal,
\begin{equation} \label{eq:T-ij-diag}
 (T_{ij}) =
 \begin{pmatrix}
  \phi_1 - \frac{\phi_5}{\sqrt{3}} & 0 & 0 \\
  0 & -\phi_1 - \frac{\phi_5}{\sqrt{3}} & 0\\
  0 & 0 & 2\frac{\phi_5}{\sqrt{3}}
 \end{pmatrix},
\end{equation}
we can write $(\phi_a) = (\phi \sin \xi, 0, 0, 0, \phi \cos \xi)$ with $\phi
\coloneqq \sqrt{\phi_a\phi_a}\geq 0$.
Shifting the parameter $\xi$ by $\xi \mapsto \xi + 2\pi/3$ corresponds to a
cyclic permutation of the $x$-, $y$-, and $z$-axes. E.g., the state $(\phi_a) =
\phi (\sqrt{3}/2, 0, 0, 0, 1/2)$ for $\xi = \pi/3$ transforms into the state
$(\phi_a) = \phi (0,0,0,0,-1)$ for $\xi = \pi$ by permuting $(x,y,z) \mapsto
(y,z,x)$. We may thus restrict the range of $\xi$ to $0 \leq \xi < 2\pi/3$.
Finding a finite $\phi \neq 0$ to be energetically favorable corresponds to a
spontaneous breaking of the rotational symmetry.
While for generic $\xi$ no continuous part of the symmetry is left intact, for
$\xi \equiv 0 \mod 2\pi/3$ or $\xi \equiv \pi/3 \mod 2\pi/3$ only two generators
of the $\mathrm O(3)$ are broken, with a residual $\mathrm O(2)$ symmetry
resulting. The corresponding \emph{uniaxial} states $(\phi_a) = (0,0,0,0,\pm
\phi)$ (modulo rotations) are characterized by a single director, in analogy to
the uniaxial nematic phase in liquid crystals~\cite{chaikin}.

The energy in the present basis reads as
\begin{align} \label{eq:mf-energy-1b}
 E_\text{MF}(\phi,\xi) & = \frac{r}{2} \phi^2 - 2 \left(g \phi\right)^{5/2} \int_0^{\frac{\Lambda}{\sqrt{g \phi}}} dx \int \frac{d \Omega}{(2\pi)^3}
 \nonumber \\ & \quad
 \times x^2 \sqrt{x^4+2 x^2 (\tilde d_1 \sin \xi + \tilde d_5 \cos \xi) + 1},
\end{align}
where we substituted $p / \sqrt{g \phi} \mapsto x$ and abbreviated the angular
integration as $\int d\Omega = \int_0^\pi d\theta \sin \theta \int_0^{2\pi}
d\varphi$. The integral becomes finite for $\Lambda / \sqrt{g \phi} \to \infty$
when we add a suitably written zero (corresponding to the parts in $E_\text{MF}$
that are constant and quadratic in $\phi$, respectively) as
\begin{multline}
 0 = - \frac{4\pi}{(2\pi)^3} \left(\frac{2}{5} \Lambda^5 + \frac{4}{5} \Lambda g^2 \phi^2 \right)
 \\
 + 2 (g \phi)^{5/2} \int_0^{\frac{\Lambda}{\sqrt{g \phi}}} dx \int \frac{d\Omega}{(2\pi)^3} \left(x^4+\frac{2}{5}\right).
\end{multline}
The mean-field energy then is (modulo irrelevant additive constants $\propto
\Lambda^5$)
\begin{equation} \label{eq:mf-energy-2}
 E_\text{MF}(\phi,\xi) = \frac{r'}{2} \phi^2
 + t(\xi) \left(g \phi\right)^{5/2}+ \mathcal O(\phi^3)
\end{equation}
with $r' = r - \frac{8}{5}\frac{4\pi \Lambda}{(2\pi)^3}g^2$ the curvature at the
origin and with the coefficient of the nonanalytic term $\propto \phi^{5/2}$ as
\begin{align}
 t(\xi) & = 2\int_0^\infty dx \int \frac{d\Omega}{(2\pi)^3} \biggl[x^4+\frac{2}{5}
 \nonumber \\ & \quad\qquad
 - x^2\sqrt{x^4+2 x^2 (\tilde d_1 \sin \xi + \tilde d_5 \cos \xi) + 1}
 \biggr]
 \nonumber \\ \label{eq:t-xi-2}
 & \simeq \frac{4\pi}{(2\pi)^3} \left[ \frac{\pi}{8} + \frac{1}{2}\left(\frac{19}{30} - \frac{\ln 3}{8} - \frac{\pi}{8} \right) \left(1-\cos 3\xi \right) \right].
\end{align}
The second line of Eq.~\eqref{eq:t-xi-2} approximates the numerical quadrature
within an error range of $\lesssim 0.5\%$ for generic $\xi$ and becomes exact
for $\xi = 0$ and $\xi = \pi/3$.
$t(\xi)$ is positive and bounded from below and above as $\frac{\pi}{8} \leq t(\xi) / \frac{4\pi}{(2\pi)^3} \leq \frac{19}{30}-\frac{\ln 3}{8}$.
The QC point at the critical coupling
\begin{equation} \label{eq:g2-r-crit}
 \left(\frac{g^2}{r}\right)_\mathrm{c} = \frac{5}{8} \frac{(2\pi)^3}{4\pi  \Lambda},
\end{equation}
when the curvature $r'$ of $E_\text{MF}(\phi,\xi)$ at $\phi = 0$ changes sign,
thus corresponds to a \emph{continuous} phase transition---in contrast to the
discontinuous (at least on the mean-field level) classical isotropic-to-nematic
transition in liquid crystals~\cite{chaikin}.
A similar such unconventional continuous phase transition has recently been
found in a model describing the spontaneous breaking of time-reversal symmetry
in the pyrochlore iridates~\cite{savary}.

\begin{figure}
 {\centering\includegraphics[height=5.5cm]{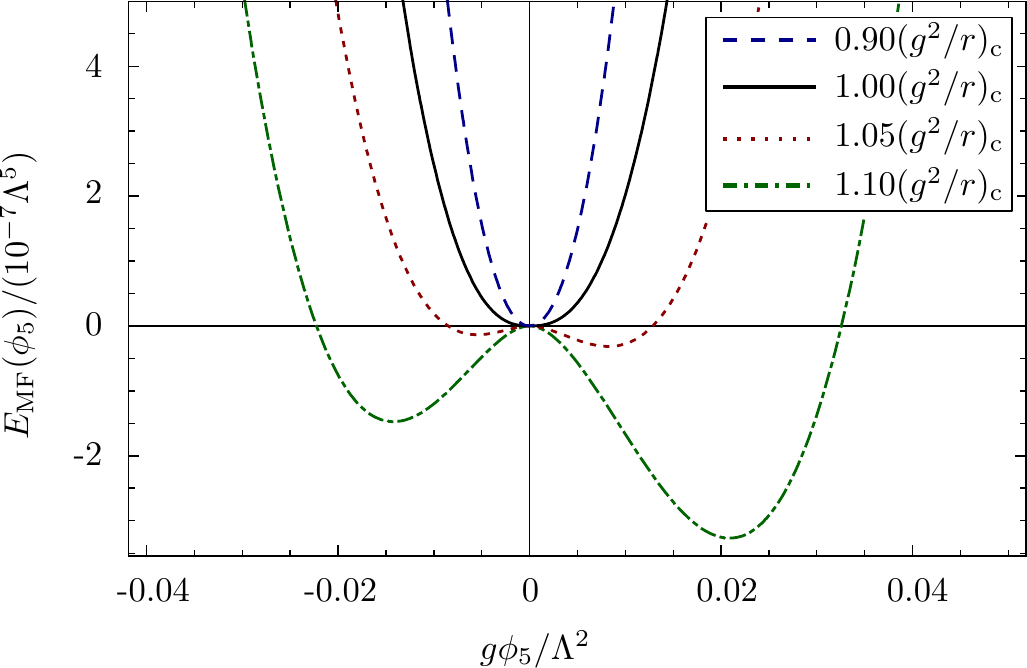}}
 \caption{Mean-field energy $E_\text{MF}(\phi_5)$ for the uniaxial states
$(\phi_a) = (0,0,0,0,\phi_5)$ that preserve a residual $\mathrm O(2)$ symmetry
(i.e., $\xi \equiv 0 \mod 2\pi / 3$ or $\xi \equiv \pi/3 \mod 2\pi/3$) for
different values of the coupling $g^2/r$ in the vicinity of the critical
coupling $(g^2/r)_\mathrm c$. The unique absolute minimum of the potential is at
zero or positive $g \phi_5$, corresponding to the isotropic state and uniaxial
nematic fully gapped state, respectively. The transition into the latter phase
for overcritical coupling is continuous.}
 \label{fig:mf-energy}
\end{figure}

$t(\xi)$ attains its unique miminum at $\xi = 0$. When $g^2/r > (g^2/r)_\mathrm
c$ the transition is thus into the state with the order parameter $(\phi_a) =
(0,0,0,0,\phi)$, $\phi > 0$, which breaks the rotational $\mathrm O(3)$ symmetry
but leaves rotations about the $z$-axis intact. The spectrum of fermions in this
state has a full, anisotropic ($\theta$-dependent) gap, with the minimal value
at $\theta = \pi/2$ and $p^2 = g \phi/2$ of $\sqrt{3} g \phi /2$. The system
appers as if under (dynamically generated) uniaxial strain~\cite{moon,
romanewald}, and, for the systems with the band structure equivalent to that of
$\alpha$-Sn or HgTe, represents a topological Mott insulator~\cite{janssen}.
We depict the mean-field energy $E_\text{MF}(\phi_5)$ for the $\mathrm
O(2)$-invariant states $(\phi_a) = (0,0,0,0,\phi_5)$ for different values of the
coupling $g^2 / r$ in Fig.~\ref{fig:mf-energy}, illustrating the continuous
nature of the transition and the energetically favored minium at $g\phi_5 >
0$~\cite{note1}.

\section{RG flow equations}
\label{sec:rg-flow}

In order to show the existence of the nematic QC point beyond the mean-field
theory we include next the effects of the bosonic fluctuations. To this end we
perform the standard Wilson's renormalization group calculation, in which both
the order parameter and the fermionic fields with the momenta within the
momentum shell $[\Lambda/b, \Lambda]$ and with all Matsubara frequencies are
integrated out~\cite{book}.
At the critical surface $r=0$, to the leading order in the self-interaction
$\lambda$ and the Yukawa coupling $g$ the result is the differential flow of the
couplings:
\begin{align}
\frac{dc}{d\ln b} & = (2 - 2 z - \eta_{\phi} ) c + \frac{2}{5} g^2 + \frac{21}{4} \sqrt{c} \lambda^2,
\label{eq:beta-c} \displaybreak[0]\\
\frac{d g}{d\ln b} & = \frac{1}{2} (6-d-z -\eta_{\phi} - 2 \eta_{\psi}) g +   \frac{6}{5} H(c) g^3, \label{eq:beta-g} \displaybreak[0]\\
\frac{d \lambda }{d\ln b} & = \frac{1}{2} (6-d-z - 3 \eta_{\phi}) \lambda  - \frac{27}{4}\frac{\lambda^3}{\sqrt{c}} - \frac{\sqrt{3}}{35}  g^3.	 \label{eq:beta-lambda}
\end{align}
Here, we have rescaled the couplings as $g^2 \Lambda^{d + z + \eta_\phi +
2\eta_\psi - 6} S_d /(2\pi)^d \mapsto g^2$ and $\lambda^2 \Lambda^{d + z +
3\eta_\phi - 6} S_d /(2\pi)^d \mapsto \lambda^2$ with $S_d$ the surface area of
the $(d-1)$-sphere.
The parameter $c$ has been rescaled as $c \Lambda^{2z + \eta_\phi - 2}
\mapsto c$.
The order parameter's and the fermion's anomalous dimensions, and the dynamical
critical exponent are to the leading order
\begin{align}
\eta_{\psi} & = \frac{4}{5} F(c) g^2, \\
\eta_{\phi} & = \frac{44}{35} g^2 + \frac{21}{4} \frac{\lambda^2}{\sqrt{c}}, \\
z & = 2 + \frac{5}{2} G(c) g^2 - \eta_{\psi}.  \label{eq:z}
\end{align}
The functions $F(c)$, $G(c)$, and $H(c)$ are the result of the one-loop
frequency integrals, and are defined as
\begin{align}
F(c) & = \frac{8 + 9 \sqrt{c} + 3 c}{8 \left(1 + \sqrt{c}\right)^3}, \label{eq:F(c)} \\
G(c) & = \frac{\sqrt{c}}{\left(1 + \sqrt{c}\right)^2}, \label{eq:G(c)} \\
H(c) & = \frac{4 + 3 \sqrt{c}}{4 \left(1 + \sqrt{c} \right)^2}. \label{eq:H(c)}
\end{align}

Small perturbations out of the critical surface are relevant in the sense of the
RG, and governed by the flow equation
\begin{equation} \label{eq:beta-r}
 \frac{d r}{d \ln b} = (2 - \eta_\phi) r - \frac{8}{5} g^2 - 21 \frac{\lambda^2}{\sqrt{c}} \frac{1}{\left(1+r\right)^{3/2}},
\end{equation}
where we have rescaled $r \Lambda^{\eta_\phi - 2} \mapsto r$.

Two comments on the computation of the RG flow equations are in order:
First, we have chosen the anomalous dimensions $\eta_\psi$ and $\eta_\phi$ and
the dynamical exponent $z$ so that the coefficients in both terms in $L_{\psi}$
as well as the momentum term in $L_\phi$ [i.e., $T_{ij} (-\nabla^2) T_{ji}/4$]
remain unity after the mode elimination, which forces the remaining coefficient
$c$ in $L_{\phi}$ then to flow.
However, while $c$ is irrelevant at the Gaussian fixed point, its stable
fixed-point value is shifted towards finite $c>0$ when $g \neq 0$. At an
interacting fixed point $c$ thus scales as $c \propto \xi^{2z + \eta_\phi - 2}$
relative to a characteristic (diverging) length scale $\xi \propto
\omega^{-1/z}$. The scaling form of the inverse two-point function at the
anticipated QC point then is
\begin{equation}
\left\langle \phi_a(\omega, \vec p)\, \phi_b(0,0) \right\rangle^{-1} = p^{2-\eta_\phi} f\left(\frac{\omega}{p^z}\right)\delta_{ab}
\end{equation}
with the scaling function $f$ that has the asymptotic limits
\begin{equation}
 f(x) \propto
 \begin{cases}
  1 & \text{for } x \ll 1, \\
  x^{(2-\eta_\phi)/z} & \text{for } x \gg 1.
 \end{cases}
\end{equation}
The alternative scaling prescription that chooses the anomalous dimensions and
the dynamical exponent such that the coefficients of both the momentum and
frequency terms in $L_\phi$ remain fixed, and in turn allows a flowing parameter
$a$ in front of the frequency term in $L_\psi$, $a (\psi^\dagger \partial_\tau
\psi)$, leads to the equivalent flow equations and same universal predictions at
the interacting fixed point (see Appendix~\ref{app:dyn-scaling}).

Second, in order to arrive at Eqs.~\eqref{eq:beta-c}--\eqref{eq:beta-r},
we have kept the general counting of dimensions in the couplings, but have
performed the angular integrations directly in $d=3$ spatial dimensions. For
details we refer to Appendix~\ref{app:RG-flow-diagrams}. In
Appendix~\ref{app:RG-four-dimensions} we present the analogous derivation of the
RG flow for the theory near $d=4$ with nine-component order-parameter field
$\phi_a$ and $16\times 16$ gamma matrices $\gamma_a$, $a=1,\dots,9$.

The mean-field result from the previous section can be recovered by neglecting
all bosonic fluctuations (e.g., by reintroducing the flavor number~$N$
and taking the limit of large $N$). The flow equation for the coupling $g^2/r$
then becomes
\begin{equation}
 \frac{d (g^2/r)}{\ln b} =
 (2-d) \frac{g^2}{r} + \frac{8}{5} \left(\frac{g^2}{r}\right)^2,
\end{equation}
which in $d=3$ has the zero exactly at the mean-field critical coupling $(g^2 /
r)_\mathrm c = 5/8$, cf.\ Eq.~\eqref{eq:g2-r-crit} and the coupling rescalings
below Eqs.~\eqref{eq:beta-lambda} and \eqref{eq:beta-r}.

To show that there exists a stable (quantum critical) fixed point of the
equations also at $N=1$ we introduce new variables
\begin{align}
  u & = \frac{\lambda}{c^{1/4} }, &
  v & = \frac{g}{c^{1/12}},
\end{align}
with $c$ chosen such that it satisfies its own fixed-point equation
\begin{equation} \label{eq:fp-eq-c}
 0 = \left(2 - 2z\right) c  + \left(\frac{2}{5} -  \frac{44}{35} c \right) c^{1/6} v^2.
\end{equation}
In terms of the new variables we can rewrite the flow equations as
\begin{align}
\frac{du}{d\ln b} & = \frac{1}{2} \left(\epsilon + 2-z - 3\eta_\phi \right) u
- \frac{27}{4} u^3 -
\frac{\sqrt{3}}{35} v^3, \\
\frac{dv}{d\ln b} & = \frac{1}{2}\left(\epsilon + 2-z - \eta_\phi - 2\eta_\psi
\right) v + \frac{6}{5} c^{\frac{1}{6}} H(c) v^3,
\end{align}
where we used Eq.~\eqref{eq:fp-eq-c} and also displayed the small
parameter $\epsilon=4-d$.

After this change of variables, the stable fixed point is readily found to
lie at $u = \mathcal O(\epsilon^{1/2})$, $v = \mathcal O(\epsilon^{1/2})$, and
$c = \mathcal O(\epsilon^{6/5})$. Since $F(0) = 1$, $G(0) = 0$, and $H(0) = 1$,
the fixed point features the critical exponents at leading order in $\epsilon$
\begin{align} \label{eq:exponents-fixed-point}
 \eta_\psi & = \mathcal O(\epsilon^{6/5}), &
 \eta_\phi & = \epsilon + \mathcal O(\epsilon^{6/5}), &
 z & = 2 + \mathcal O(\epsilon^{6/5}),
\end{align}
and it is located at the values of $u$ and $v$ that satisfy the equations:
\begin{align}
\frac{21}{4} u^2 & = \epsilon, &
\frac{\sqrt{3}}{35} v^3 & = u \left( -\epsilon - \frac{27}{4} u^2 \right).
\end{align}
The last equation, in particular, implies that at the fixed point
$\sign(v)=-\sign(u)$, whereas the first one leaves the sign of $u$ undetermined.
We find the following finite fixed-point values
\begin{align}
 u^*_\pm & = \mp \frac{2}{\sqrt{21}} \sqrt{\epsilon}, &
 v^*_\pm & = \pm 2 \left(\frac{20}{3}\right)^{\frac{1}{3}} \left(\frac{1}{7}\right)^{\frac{1}{6}} \sqrt{\epsilon}.
\end{align}
As the partition function is invariant under the \emph{simultaneous} sign
change of $g$ and $\lambda$, so are the flow equations. Thus, the two fixed
points at $(u^*_\pm, v^*_\pm)$ are physically equivalent. It is easy to check
that this fixed point is indeed critical, i.e., with no other unstable
directions except for the direction of the tuning parameter~$r$. From the flow
of $r$ we find the exponent $\nu$ that governs the scaling of the correlation
length $\xi \propto |\delta|^{-\nu}$, with $\delta$ denoting the deviation from
the critical point, as
\begin{equation} \label{eq:1-nu}
 1/\nu = 2 + 5 \epsilon + \mathcal O(\epsilon^{6/5}).
\end{equation}
Notably, the correction to the mean-field exponent $1/\nu = 2$ is positive, in
contrast to the QC points in Dirac fermion systems that are described by the
$z=1$ Gross-Neveu universality classes~\cite{herjurvaf, janssenherbut}.
The reason for the difference in sign is the presence of the cubic term
$\propto \Tr T^3$ in the action, which renders the bosonic contribution to the
flow equation of the tuning parameter~$r$ [last term in Eq.~\eqref{eq:beta-r}]
of opposite sign as compared to systems with quartic bosonic interactions.
Likewise, in the field theory of the classical isotropic-to-nematic phase
transition in liquid crystals which allows the cubic tensor invariant the leading
correction to $1/\nu$ is also positive~\cite{lubenskypriest}.

We emphasize that our quantitative predictions for the critical exponents
obtained near the upper critical dimension, although interesting
in their own right, may not describe well real 3D
systems in which long-range Coulomb interaction is important, such as
$\alpha$-Sn or HgTe. In these cases, a nematic gap might already open up at
infinitesimal coupling~\cite{janssen}. Our result, however, at the very least shows that a QC
fixed point \emph{exists} near the upper critical dimension, substantiating the
``fixed-point-collision'' scenario, and it allows to study the
\emph{qualitative} properties of the nematic instability. We
therefore refrained from displaying the subleading terms $\propto
\epsilon^{6/5}$ in Eqs.~\eqref{eq:exponents-fixed-point} and \eqref{eq:1-nu},
which are  straightforwardly computable from our one-loop flow equations,
but do not, in our opinion, necessarily have direct relevance for the physics in $d=3$.

\begin{figure}
 {\centering \includegraphics[width=0.48\textwidth]{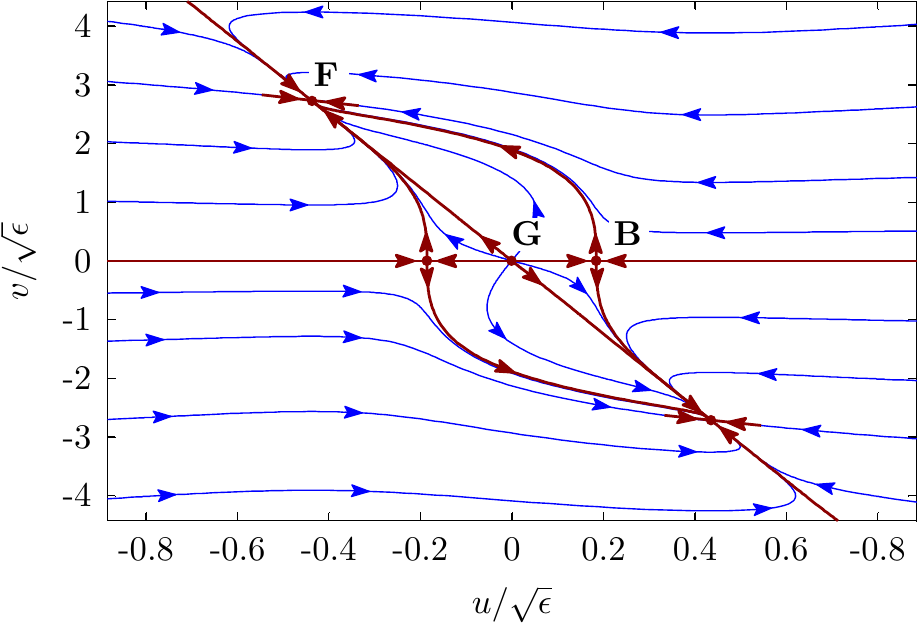}}
 \caption{RG flow in the $u$-$v$ plane for $r=0$ and $c=\mathcal
 O(\epsilon^{6/5})$, to leading order in $\epsilon$. Arrows point towards
 infrared. The purely bosonic fixed point (B) is unstable in direction of the
 Yukawa coupling $v$. The fermionic fixed point (F) is critical, with no other
 unstable directions except for the direction of the tuning parameter~$r$. It
 governs the transition into an infrared phase that has fully gapped fermions
 and a spontaneously broken rotational symmetry.}
 \label{fig:rg-flow}
\end{figure}

We have plotted the leading-order RG flow in the $u$-$v$ plane for $c =
\mathcal O(\epsilon^{6/5})$ in Fig.~\ref{fig:rg-flow}, showing besides the
unstable Gaussian~(G) and stable fermionic~(F) fixed points also the purely
bosonic fixed point~(B) at $v=0$ and finite $u \neq 0$. B is unstable in the
direction of $v$, in analogy to the bosonic Wilson-Fisher fixed point in Dirac
fermion systems~\cite{herjurvaf}.

In the calculation with four-dimensional tensor order parameter we find that
the bosonic fixed point B disappears, in full analogy to the $p=4$ critical
point in the field theory of the classical isotropic-to-nematic phase transition
in liquid crystals~\cite{lubenskypriest}. In contrast, the fermionic fixed point
(F in Fig.~\ref{fig:rg-flow}) survives for any dimension $p$ of the tensor
field, with changing only its stability properties at larger values of $p$.
Furthermore, our universal predictions for the anomalous dimensions
$\eta_\phi$ and $\eta_\psi$ as well as the critical exponents $z$ and $1/\nu$ at
the fermionic fixed point turn out to agree at leading order exactly with
Eqs.~\eqref{eq:exponents-fixed-point} and \eqref{eq:1-nu}, see
Appendix~\ref{app:RG-four-dimensions}.

\section{Interpretation}
\label{sec:interpretation}

At the mean-field level, the model features a continuous (quantum) phase
transition, described by the large-$N$ fixed point of the Gross-Neveu-Yukawa
field theory. It therefore seems natural to associate the identified fixed point
also for $N=1$ with a continuous nematic quantum phase transition. One should
note, however, that near the upper critical dimension and at small $N$ the
parameter $c$ in $L_\phi$ becomes small at the fixed point, emphasizing the
significance of purely bosonic fluctuations. The result of the mean-field theory
may thus as well be overturned  in the physical limit, and the possibility that
at small $N$ the nature of the transition differs from the mean-field picture
cannot be excluded with certainty. We believe, nonetheless, that even in a
scenario with a discontinuous quantum phase transition the above critical fixed
point would still retain its physical significance: such a situation arises, for
example, in the related classical Ginzburg-Landau-Wilson theory for the
(presumably discontinuous) thermal isotropic-to-nematic transition in liquid
crystals, which also exhibits a critical fixed point in the related
$\epsilon=6-d$ expansion~\cite{lubenskypriest}. A plausible interpretation of
the latter is that it describes the disappearance of the energy barrier between
the high- and low-temperature phases, and the ultimate instability of the
metastable symmetric phase.

An interesting feature of the identified fixed point is worth pointing out.
In the reference frame where the nematic tensor would become diagonal
[Eq.~\eqref{eq:T-ij-diag}], the bosonic part of the Lagrangian for uniform order
parameter $(\phi_a) = (\phi \sin \xi,0,0,0,\phi \cos\xi)$, $\phi > 0$, becomes
\begin{equation}
  L_{\phi} = \frac{r}{2} \phi^2
  + \frac{2\lambda}{\sqrt{3}} \cos(3\xi) \phi^3
  + b \phi^4 + \mathcal O(\phi^5),
\end{equation}
where we have displayed the unique symmetry-allowed quartic term as well.
At intermediate steps of the RG the Lagrangian is analytic in $\phi$, and the
nonanalytic term $\propto \phi^{5/2}$ will only emerge in the deep infrared,
when all modes are integrated out. During this process it may in general receive
contributions from the flow of all higher-order terms. If we focus for
simplicity only on the leading cubic and quartic invariants and consider
(without loss of generality) the fixed point at $\lambda < 0$, we find
that the effective quantum potential at the fixed point is minimized for $\xi =
0$.
If this remains true up to the infrared, when the $\phi^{5/2}$ term in $L_\phi$
emerges, it indicates that the interacting ground state for strong coupling has
(also for small $N$) the uniaxial form with $\phi_5 > 0$ and $\phi_1 = 0$.
The fate of fermions in this state depends crucially then on the sign of the
remaining Yukawa coupling $g$. If $g > 0$, the combination $g \phi_5 > 0$, and
we recover the mean-field ground state with the spectrum of fermions having the
full, anisotropic gap (cf.\ Sec.~\ref{sec:mf-theory}). If, on the other hand,
$g < 0$ and  $g \phi_{5}<0$, the spectrum has two gapless points in the vicinity
of which the energy dispersion becomes linear~\cite{janssen}.

We see however, that the leading term in the flow equation for $\lambda$ is
$-g^3$, so that a negative self-interaction $\lambda$ is generated only by a
{\it positive} Yukawa coupling $g$. This is reflected in the fixed-point
location, at which the signs of the two couplings are inevitably opposite.
Incidentally, this feature is also responsible for the stability of the fixed
point. Also, even if we start the RG flow at microscopic couplings $\lambda$ and
$g$ of the same sign, we always flow to a regime in which $\lambda g<0$, at
least in the vicinity of the critical surface (see Fig.~\ref{fig:rg-flow}). We
thus find that the consistent theory in the infrared has the fermions fully
gapped in the broken symmetry phase, in agreement with the mean-field result.

\section{Conclusions}
\label{sec:conclusions}

In sum, we constructed the field theory of the fermions with the chemical
potential at the point of quadratic band touching in three spatial dimensions
coupled to the second-rank tensorial nematic order parameter. We argued that
this field theory has an upper critical dimension of four, and that it possesses
a perturbatively accessible quantum critical point in the vicinity of four
dimensions.
The critical point governs the (presumably continuous) transition between the
semimetal to the fully, but anisotropically gapped Mott insulator.
The existence of the critical point in the theory supports the scenario of the
``fixed-point collision''~\cite{janssen},
according to which the Fermi system with QBT in presence of the long-range
tail of Coulomb interaction, which we have here suppressed, features a lower
critical dimension $d_\text{low}$ with $2 < d_\text{low} < 4$. At $d_\text{low}$
the nematic QC point and the Abrikosov's NFL fixed point collide and then
disappear from the real space of physical couplings, leaving behind the runaway
flow. The ground-state physics of the 3D systems such as clean $\alpha$-Sn or HgTe
crucially depends on whether $d_\text{low}$ is above or below $d=3$. The
one-loop analysis points to $d_\text{low}$ slightly above three, which would
make the QBT point unstable towards the nematic insulator even in the
weak-coupling limit~\cite{janssen}.
If, on the other hand, the true value of $d_\text{low}$ would turn out to be
below three and both the Abrikosov's NFL as well as the QC fixed points persist
all the way down to the physical dimension, the weakly-interacting systems are
governed by the attractive NFL fixed point and should exhibit anomalous power
laws in several observables~\cite{moon}.
If $d_\text{low}$ is below but not too far from $d=3$, however, and the nematic
QC point is consequently located at not too large a coupling, one could still
speculate on situations, e.g., in \emph{uniformly} strained systems or in cold-atom
quantum simulators, in which the interactions may be tuned through the nematic
QC point. In any case, it would obviously be desirable to gain a firmer theoretical
control over the true value of $d_\text{low}$. This work is under
way~\cite{janssen2}.

\acknowledgments{We thank F.~Assaad, B.~D\'{o}ra, H.~Gies, L.~Golubovi\'{c},
Y.~Meurice, W.~Metzner, A.~Rosch, B.~Roy, B.~Skinner,  and especially
Oskar Vafek for discussions on this and related subjects, and we are
grateful to M.~Vojta for pointing us to  Ref.~\cite{meng} and the problem of
multiple dynamic scaling. The support from the DFG under JA\,2306/1-1 and the
NSERC of Canada is acknowledged.}

\appendix

\section{Generalized real Gell-Mann matrices}
\label{app:Gell-Mann-matrices}

For completeness, let us review the construction of the generalized Gell-Mann
matrices in $d$ dimensions (the generators of $SU(d)$)~\cite{hioeeberly}. They
can be classified into three groups.
The first group is given by the real, diagonal, and traceless matrices
\begin{equation}
 \hat w_{l} = - \sqrt{\frac{2}{l(l+1)}} \sum_{j=1}^{l} \left(|j \rangle \langle j | - |l+1 \rangle \langle l+1| \right),
\end{equation}
where $1 \leq l \leq d-1$ and $|1 \rangle, \dots, |d \rangle$ denote the
(standard) orthonormal basis vectors in $\mathbbm R^d$, $\langle i | j \rangle =
\delta_{ij}$.
The second group are $d(d-1)/2$ real symmetric matrices that have nonvanishing
elements only on the off-diagonal, namely the matrices $\hat u_{jk}$ with ones
in the $jk$-th and $kj$-th entries and zero otherwise
\begin{equation}
 \hat u_{jk} = |j \rangle \langle k | + |k \rangle \langle j |, \qquad \text{where } 1\leq j < k \leq d.
\end{equation}
The third group are $d(d-1)/2$ imaginary matrices which can be constructed
similarly to $\hat u_{jk}$. However, for the purposes of the present work we
only need the \emph{real} Gell-Mann matrices of the first and second group.

In $d=2$, this construction gives $\hat w_1 = - \sigma_3$ and $\hat u_{12} =
\sigma_1$.
In $d=3$, we recover the standard (modulo name and sign conventions) $3 \times
3$ real Gell-Mann matrices:
\begin{gather}
 \Lambda_1 = -\hat w_1 =
 \begin{pmatrix}
  1 & 0 & 0 \\
  0 & -1 & 0 \\
  0 & 0 & 0
 \end{pmatrix}, \qquad
 \Lambda_2 = \hat u_{12} =
 \begin{pmatrix}
  0 & 1 & 0 \\
  1 & 0 & 0\\
  0 & 0 & 0
 \end{pmatrix}, \displaybreak[0] \nonumber\\
 \Lambda_3 = \hat u_{13} =
 \begin{pmatrix}
  0 & 0 & 1 \\
  0 & 0 & 0 \\
  1 & 0 & 0
 \end{pmatrix}, \qquad
 \Lambda_4 = \hat u_{23} =
 \begin{pmatrix}
  0 & 0 & 0 \\
  0 & 0 & 1 \\
  0 & 1 & 0
 \end{pmatrix}, \displaybreak[0] \nonumber\\
 \Lambda_5 = \hat w_2 =
 \frac{1}{\sqrt{3}}
 \begin{pmatrix}
  -1 & 0 & 0\\
  0 & -1 & 0\\
  0 & 0 & 2
 \end{pmatrix}.
\end{gather}
In $d=4$, we find
\begin{gather}
 \Lambda_1 =
 \begin{pmatrix}
  1 & 0 & 0 & 0 \\
  0 & -1 & 0 & 0 \\
  0 & 0 & 0 & 0 \\
  0 & 0 & 0 & 0
 \end{pmatrix}, \qquad
 \Lambda_2 =
 \begin{pmatrix}
  0 & 1 & 0 & 0 \\
  1 & 0 & 0 & 0 \\
  0 & 0 & 0 & 0 \\
  0 & 0 & 0 & 0
 \end{pmatrix}, \displaybreak[0] \nonumber\\
 \Lambda_3 =
 \begin{pmatrix}
  0 & 0 & 1 & 0 \\
  0 & 0 & 0 & 0 \\
  1 & 0 & 0 & 0 \\
  0 & 0 & 0 & 0
 \end{pmatrix}, \qquad
 \Lambda_4 =
 \begin{pmatrix}
  0 & 0 & 0 & 0 \\
  0 & 0 & 1 & 0 \\
  0 & 1 & 0 & 0 \\
  0 & 0 & 0 & 0
 \end{pmatrix}, \displaybreak[0] \nonumber\\
 \Lambda_5 =
 \frac{1}{\sqrt{3}}
 \begin{pmatrix}
  -1 & 0 & 0 & 0 \\
  0 & -1 & 0 & 0 \\
  0 & 0 & 2 & 0 \\
  0 & 0 & 0 & 0
 \end{pmatrix},  \qquad
 \Lambda_6 =
 \begin{pmatrix}
  0 & 0 & 0 & 1 \\
  0 & 0 & 0 & 0 \\
  0 & 0 & 0 & 0 \\
  1 & 0 & 0 & 0 \\
 \end{pmatrix}, \displaybreak[0] \nonumber\\
 \Lambda_7 =
 \begin{pmatrix}
  0 & 0 & 0 & 0  \\
  0 & 0 & 0 & 1 \\
  0 & 0 & 0 & 0 \\
  0 & 1 & 0 & 0
 \end{pmatrix}, \qquad
 \Lambda_8 =
 \begin{pmatrix}
  0 & 0 & 0 & 0  \\
  0 & 0 & 0 & 0 \\
  0 & 0 & 0 & 1 \\
  0 & 0 & 1 & 0 \\
 \end{pmatrix}, \nonumber\\
 \Lambda_9 =
 \frac{1}{\sqrt{6}}
 \begin{pmatrix}
  -1 & 0 & 0 & 0 \\
  0 & -1 & 0 & 0 \\
  0 & 0 & -1 & 0 \\
  0 & 0 & 0 & 3
 \end{pmatrix}.
\end{gather}

In general dimension $d$, the $(d^2-d)/2 + (d-1)$ off-diagonal and diagonal,
respectively, real matrices $\Lambda_a$ form an orthogonal set:
\begin{equation}
\Tr (\Lambda_a \Lambda_b) = 2\delta_{ab},
\end{equation}
and together with the unit matrix, they form a basis in the space of real
symmetric $d$-dimensional matrices. We can therefore write the matrix element of
any symmetric matrix $M$ as
\begin{equation}
 M_{ij} = \frac{1}{d} \delta_{ij} M_{kk} + \frac{1}{2} M_{lm}\Lambda_{a,ml} \Lambda_{a,ij}
 \end{equation}
 or equivalently, as
 \begin{multline}
 \frac{1}{2} \left( \delta_{li}\delta_{mj}+ \delta_{lj}\delta_{mi} \right) M_{lm} \\ =
 \left(\frac{1}{d} \delta_{ij} \delta_{lm} + \frac{1}{2}\Lambda_{a,ml} \Lambda_{a,ij}\right) M_{lm}.
 \end{multline}
From here we deduce an important relation:
\begin{equation} \label{eq:Lambda-Lambda}
\Lambda_{a,ml} \Lambda_{a,ij}= \delta_{li}\delta_{mj}+ \delta_{lj}\delta_{mi}- \frac{2}{d}  \delta_{ij} \delta_{lm},
\end{equation}
which we use in the computation of the RG flow equations.

\section{QBT Hamiltonian in $d$ dimensions}
\label{app:QBT-general}

We can construct the general QBT Hamiltonian $H = G_{ij} p_i p_j$ in $d$
dimensions with the help of $(d^2-d)/2+(d-1) = (d+2)(d-1)/2$ gamma matrices
$\gamma_a$. They have dimension $d_\gamma = 2^{\lfloor(d+2)(d-1)/4\rfloor}$ with
$\lfloor\, \cdot \,\rfloor$ denoting the floor function. The relationship
between
the $G_{ij}$ and the gamma matrices $\gamma_a$, $a = 1, \dots, (d+2)(d-1)/2$ are
given by the real and symmetric (generalized) $d \times d$ Gell-Mann matrices
$\Lambda_a$ as
\begin{equation}
 G_{ij} = \sqrt{\frac{d}{2(d-1)}} \Lambda_{a,ij} \gamma_a.
\end{equation}
Together with the Clifford algebra $\{ \gamma_a, \gamma_b \} = 2 \delta_{ab}$
and Eq.~\eqref{eq:Lambda-Lambda}, this immediately gives $H^2 = p^4$, as
expected. In any dimension, we can thus write the Hamiltonian in the form
\begin{equation}
H = d_a(\vec p) \gamma_a , \qquad a = 1, \dots, \tfrac{1}{2}(d+2)(d-1),
\end{equation}
with
\begin{equation} \label{eq:da-Lambda}
 d_a (\vec p) = p^2 \tilde d_a (\Omega) = \sqrt{\frac{d}{2(d-1)}} p_i \Lambda_{a,ij} p_j.
\end{equation}
This defines the real hyperspherical harmonics $\tilde d_a(\Omega)$ for angular
momentum of two in general dimension, with $\Omega$ denoting the spherical
angles on the $(d-1)$-sphere in $\vec p$-space.

\section{Computation of RG flow equations}
\label{app:RG-flow-diagrams}

\begin{figure}
 {\centering\includegraphics[width=.48\textwidth]{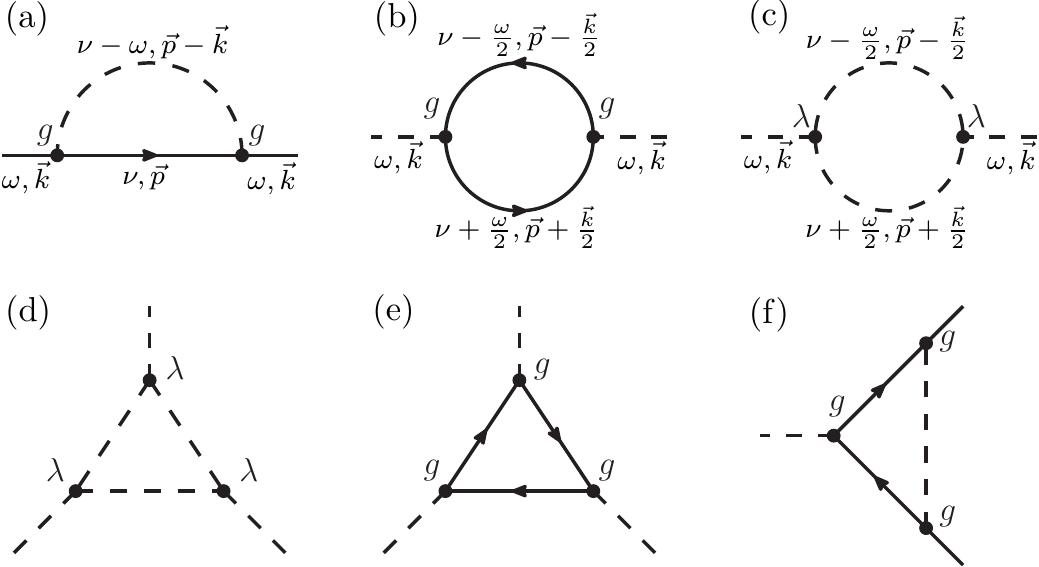}}
 \caption{Diagrams that contribute to the flow equations. Solid lines represent
fermions and dashed lines represent bosons. Top: contributions to (a) $\eta_1$,
$\eta_\psi$ and (b) $\eta_\phi$, $r$, $\beta$ and (c) $\eta_\phi$, $r$.
At finite external momenta $\vec k$ and frequencies
$\omega$ the diagrams are evaluated using the momentum and frequency routings
as displayed, with loop momentum $|\vec p|\in[\Lambda/b,\Lambda]$ and
frequency $\nu \in (-\infty,\infty)$.
Bottom: contributions to (d), (e) $d\lambda / d \ln b$ and (f) $d g / d \ln b$.}
 \label{fig:diagrams}
\end{figure}

Let us provide some details on the computation of the RG flow
equations~\eqref{eq:beta-c}--\eqref{eq:z}. In the perturbative expansion, after
integrating out the high-energy modes with momenta in the thin shell
$[\Lambda/b, \Lambda]$, we arrive at the effective action for the low-energy
modes
\begin{align} \label{eq:low-energy-action}
 S_< & = \int_0^{\Lambda/b} \frac{d \vec k}{(2\pi)^d} \int_{-\infty}^\infty \frac{d \omega}{2\pi}
 \biggl[ \psi^\dagger \left( b^{\eta_1} i \omega + b^{\eta_\psi} d_a(\vec k) \gamma_a \right) \psi
 \nonumber \\ & \quad \qquad
 + \frac{1}{2} \phi_a \left( \left(c+\delta c\right) \omega^2 + b^{\eta_\phi}
 k^2  + \left(r+\delta r\right) \right) \phi_a
 \nonumber \\ & \quad \qquad
 + \left(\beta + \delta \beta\right) k_i k_j \Lambda_{a,il}\Lambda_{b,lj}
\phi_a \phi_b \biggr]
 \nonumber \\ & \quad
 + \int_0^{\Lambda/b} \frac{d \vec k_1 d \vec k_2}{(2\pi)^{2d}} \int_{-\infty}^\infty \frac{d \omega_1 d \omega_2}{(2\pi)^2}
 \Bigl[ \left(g + \delta g \right) \left(\phi_a \psi^\dagger \gamma_a\psi\right)
 \nonumber \\ & \quad \qquad
 + \left(\lambda + \delta \lambda \right) \Lambda_{a,ij} \Lambda_{b,jl} \Lambda_{c,li} \left(\phi_a \phi_b \phi_c \right) \Bigr],
\end{align}
where we have included the second symmetry allowed, quadratic, momentum-dependent
term $\propto \beta k_i k_j T_{il}
T_{lj}$ (third line in Eq.~\eqref{eq:low-energy-action}) for generality.
The anomalous dimensions $\eta_1$, $\eta_\psi$, and $\eta_\phi$ and the
explicit renormalizations $\delta c$, $\delta r$, $\delta\beta$,
$\delta g$, and $\delta
\lambda$ are determined by evaluating the corresponding one-loop diagrams, as
depicted in Fig.~\ref{fig:diagrams}.
$\eta_1$ and $\eta_\psi$ are given by the
fermion-boson loop in Fig.~\ref{fig:diagrams}\,(a), expanded to first order in
external frequency $\omega$ and second order in external momentum~$k$,
respectively. $\eta_\phi$ has two contributions, given by the diagrams in
Fig.~\ref{fig:diagrams}\,(b,\,c), when expanded to second order in external
momentum.
When alternatively expanded in frequency, these diagrams
explicitly renormalize the frequency term $\propto c \omega^2 \phi^2_a$.
The constant parts of the diagrams determine the shift of the tuning
parameter $r$.
At the upper critical dimension the coefficients of the diagrams become
universal (see Appendix~\ref{app:RG-four-dimensions}). Here, we
will perform the angular integrations directly in $d=3$ spatial dimensions. The
diagrams may then receive (slight) regularization dependencies. To be explicit,
we distribute finite external momentum and frequency in the fermion-boson loop
in Fig.~\ref{fig:diagrams}\,(a) such that the fermion loop momentum is
on-shell,
$|\vec p| \in [\Lambda/b,\Lambda]$, while we choose a symmetric momentum and
frequency distribution in the fermion-fermion and the boson-boson loops
in Fig.~\ref{fig:diagrams}\,(b,\,c).

One further comment on the bosonic contribution to $\eta_\phi$
[Fig.~\ref{fig:diagrams}\,(c)] should be made: We note that the diagram
is invariant under the ``pseudo-relativistic'' rotation $(\sqrt c \omega, \vec
k)_\mu \mapsto O_{\mu\nu} (\sqrt c \omega, \vec k)_\mu$ with
$(d+1)$-dimensional rotation matrix $O^\mathrm T O = \mathbbm 1$, $\mu,\nu =
0,\dots,d$.
In order to compute the contribution to $\eta_\phi$ one may therefore expand
the diagram either in external frequency
$c \omega^2$ or in external momentum $k^2$, and both
prescriptions should give the same result due to the
``pseudo-relativistic'' invariance of the diagram.
Put differently, in $d c/d \ln b$, the two contributions from $\eta_\phi$ and
$\delta c$ from this diagram should cancel.
The invariance, however,
is broken by our regularization scheme, in which we integrate out all
frequencies at once,
rendering the coefficients of the $c\omega^2$-term and the $k^2$-term
different. It is therefore \emph{a priori} not clear
which one to choose as the one giving the
contribution to $\eta_\phi$. To resolve this issue we recompute the
diagram Fig.~\ref{fig:diagrams}\,(c) using a ``pseudo-relativistic''
regularization with loop momentum and frequency $\Lambda/b \leq \sqrt{c \nu^2 +
p^2} \leq \Lambda$, which gives the same contribution independent of whether
one expands in $c\omega^2$ or $k^2$. We find that the value obtained in
this scheme is in fact exactly the  same as the one obtained by
expanding the diagram in $c\omega^2$ in our standard scheme,
so we thus use this value as the bosonic contribution to $\eta_\phi$.

The boson- and fermion-loop diagrams in Fig.~\ref{fig:diagrams}\,(d,\,e),
respectively, renormalize the bosonic self-interaction $\lambda$. In order to
evaluate these diagrams, we continually make use of the identities in
Eqs.~\eqref{eq:Lambda-Lambda} and \eqref{eq:da-Lambda} derived above. For
instance, for the evaluation of the fermion loop in Fig.~\ref{fig:diagrams}\,(e)
we need the following angular integral over the $(d-1)$-sphere in $\vec p$-space
\begin{align}
 & \int d \Omega \, d_a(\vec p) d_b(\vec p) d_c(\vec p) \nonumber \\
 & = \left(\frac{d}{2(d-1)}\right)^{3/2} \int d\Omega\, p_i p_j p_k p_l p_m p_n \Lambda_{a,ij} \Lambda_{b,kl} \Lambda_{c,mn} \nonumber \\
 & =\sqrt{\frac{d}{2(d-1)}} \frac{4 S_d}{(d-1)(d+2)(d+4)} \Tr(\Lambda_a \Lambda_b \Lambda_c) \, p^6,
\end{align}
where $S_d = 2 \pi^{d/2}/\Gamma(d/2)$ is the surface area of the $(d-1)$-sphere
and $\Omega$ again denotes the spherical angles on the sphere. The evaluation of
the triangle diagram in Fig.~\ref{fig:diagrams}\,(f), which renormalizes the
Yukawa vertex $g$, is straightforward when making use of the orthogonality of
the real spherical harmonics
\begin{align}
 \int d \Omega\, d_a(\vec p) d_b(\vec p) = \frac{2S_d}{(d+2)(d-1)} p^4 \delta_{ab}.
\end{align}

In order to bring the cutoff in $S_<$ back to $\Lambda$ we shift the momenta $b
\vec k \mapsto \vec k$ and frequencies $b^z \omega \mapsto \omega$ with suitable
dynamical exponent $z$. The coefficients of the momentum terms $\propto k^2$ in
the fermionic and bosonic propagators in the first and second line of
Eq.~\eqref{eq:low-energy-action}, respectively, can be fixed to one if we
renormalize the fields as
\begin{align} \label{eq:field-renormalizations}
 b^{-(2+d+z-\eta_\psi)/2} \psi & \mapsto \psi, &
 b^{-(2+d+z-\eta_\phi)/2}\phi & \mapsto \phi.
\end{align}
However, then only one of the frequency terms can be fixed. We choose the
fermionic term $\propto i \omega$, which is done by setting
\begin{equation} \label{eq:z-scaling1}
z=2+\eta_1 - \eta_\psi.
\end{equation}
At the noninteracting fixed point we thus have $z=2$.
The low-energy action $S_<$ is hence brought back into the same form as before
integrating out the momentum shell if the couplings are renormalized as
\begin{align}
 \frac{d c}{d \ln b} & = (2 - 2z - \eta_\phi )c + \frac{\partial \delta
c}{\partial \ln b}, \displaybreak[0] \\
 \frac{d g}{d \ln b} & = \frac{1}{2} (6 - d - z - \eta_\phi - 2 \eta_\psi) g + \frac{\partial\delta g}{\partial \ln b}, \displaybreak[0] \\
 \frac{d \lambda}{d \ln b} & = \frac{1}{2}(6 - d - z - 3\eta_\phi) \lambda + \frac{\partial \delta \lambda}{\partial \ln b}, \displaybreak[0] \\
 \frac{d r}{d \ln b} & = (2 - \eta_\phi) r + \frac{\partial \delta r}{\partial \ln b}.
\end{align}
If we rescale the parameters as $c \Lambda^{2z + \eta_\phi - 2} \mapsto c$,
$g^2 \Lambda^{d + z + \eta_\phi + 2\eta_\psi - 6}
S_d /(2\pi)^d \mapsto g^2$, $\lambda^2 \Lambda^{d + z + 3\eta_\phi - 6} S_d
/(2\pi)^d \mapsto \lambda^2$, and $r \Lambda^{\eta_\phi - 2} \mapsto r$, the
explicit evaluation of the diagrams leads to
Eqs.~\eqref{eq:beta-c}--\eqref{eq:beta-r} in the main text.

Let us comment on the $\beta$-term proportional to
\begin{equation}
 k_i k_j \Lambda_{a,il}\Lambda_{b,lj} \phi_a \phi_b = k_i k_j T_{il} T_{lj},
\end{equation}
which couples the internal rotations of the tensor $T$ to the spatial
rotations. Evaluating the particle-hole diagram in Fig.~\ref{fig:diagrams}\,(b)
for zero external frequency involves the integral
\begin{align}
  I_{ab}(\vec k) & = \int_{\Lambda/b}^\Lambda \frac{d \vec p}{(2\pi)^d} \int_{-\infty}^\infty \frac{d \nu}{2\pi}
  \nonumber \\ & \quad \qquad
  \times \Tr \Biggl[ \gamma_a \frac{i \nu + d_c(\vec p + \vec k) \gamma_c}{\nu^2 + (\vec p + \vec k)^4} \gamma_b \frac{i \nu + d_e(\vec p) \gamma_e}{\nu^2 +  p^4}\Biggr]
  \nonumber \\ &
  = \frac{S_3}{(2\pi)^3} \biggl[ - \frac{8}{5} \delta_{ab} \Lambda^2
   + \frac{44}{35} k^2 \delta_{ab}
  \nonumber \\ & \quad \qquad
   - \frac{27}{70} k_i k_j \Lambda_{a,il}\Lambda_{b,lj} \biggr]
\Lambda^{d-4} \ln b + \mathcal O(k^4),
  \label{eq:particle-hole-integral}
\end{align}
where in the last line we have for explicitness evaluated the angular integral
in $d=3$.
We note that the contribution to the bosonic propagator $\propto k^2
\delta_{ab}$ [second term in Eq.~\eqref{eq:particle-hole-integral}] is larger
than the contribution to the $\beta$-term $\propto (\Lambda_a \vec k) \cdot
(\Lambda_b \vec k)$ (third term). The anomalous dimension of the latter thus
becomes \emph{negative} and $\beta$ is irrelevant in the sense of the RG. This
justifies its omission in $L_\phi$, as anticipated in
Sec.~\ref{sec:GNY-field-theory}. Another way to view this is to regard $\beta$
as a coupling which flows according to
\begin{equation}
 \frac{d \beta}{d \ln b} = -
 \frac{44}{35} g^2 \beta - \frac{27}{140} g^2,
\end{equation}
which indeed has a stable fixed point at $\beta = -
27/176$. We note that the action is bounded from below when
\begin{equation}
 \frac{\delta^2 S}{\delta \phi^a \delta \phi^b} > 0 \qquad \Leftrightarrow \qquad \frac{1}{2} + \frac{4}{3} \beta > 0,
\end{equation}
and a negative fixed-point value for $\beta$ is still
consistent with stability.

\section{Alternative dynamical scaling}
\label{app:dyn-scaling}

We now show that the alternative dynamical scaling scheme in which we fix
the coefficient $c$ in front of the frequency term in $L_\phi$ and in turn
allow for a flowing parameter $a$ in front of the fermionic frequency term leads
to the equivalent flow equations and the same universal observables at
criticality. After integrating out the high-energy modes the low-energy
effective action can be written as
\begin{align} \label{eq:low-energy-action-2}
  &S_< = \nonumber \\
&
\int_0^{\Lambda/b} \frac{d \vec k}{(2\pi)^d} \int_{-\infty}^\infty
\frac{d \omega}{2\pi}
 \biggl[
 %
 %
\psi^\dagger \left( \left(a + \delta a\right) i \omega +
b^{\eta_\psi} d_a(\vec k) \gamma_a \right) \psi
 \nonumber \\ & \quad \qquad
 + \frac{1}{2} \phi_a \left( b^{\eta_2} \omega^2 + b^{\eta_\phi}
 k^2  + \left(r+\delta r\right) \right) \phi_a
 \nonumber \\ & \quad \qquad
 + \left(\beta + \delta \beta\right) k_i k_j \Lambda_{a,il}\Lambda_{b,lj}
\phi_a \phi_b \biggr]
 \nonumber \\ & \quad
 + \int_0^{\Lambda/b} \frac{d \vec k_1 d \vec k_2}{(2\pi)^{2d}}
\int_{-\infty}^\infty \frac{d \omega_1 d \omega_2}{(2\pi)^2}
 \Bigl[ \left(g + \delta g \right) \left(\phi_a \psi^\dagger \gamma_a\psi\right)
 \nonumber \\ & \quad \qquad
 + \left(\lambda + \delta \lambda \right) \Lambda_{a,ij} \Lambda_{b,jl}
\Lambda_{c,li} \left(\phi_a \phi_b \phi_c \right) \Bigr],
\end{align}
which is equivalent to Eq.~\eqref{eq:low-energy-action} upon identification
\begin{align}
 \eta_1 & = \frac{1}{a} \frac{\partial \delta a}{\partial \ln b}, &
 \eta_2 & = \frac{1}{c} \frac{\partial \delta c}{\partial \ln b},
\end{align}
and $c = 1/a^2$. Note that although $\omega$ and $k$ now have the same
units, the engineering dimensions of $g$ and $\lambda$ still retain their above
form, Eq.~\eqref{eq:dim-g-lambda}, albeit with a different dynamical
exponent at the Gaussian fixed point. The engineering dimension of the parameter
$a$ is
\begin{align}
 \dim[a] = 2 - z.
\end{align}
After the RG step $b k \mapsto k$, $b^z \omega \mapsto \omega$ and
renormalizations of the fields as in Eq.~\eqref{eq:field-renormalizations},
both the momentum term and the frequency term of the bosonic field in $S_<$ can
be brought back into the form of the initial action if we choose
\begin{equation} \label{eq:z-scaling2}
 z = \frac{1}{2} (2 + \eta_2 - \eta_\phi),
\end{equation}
and thus $z=1$ in the noninteracting limit. This is in contrast to
Eq.~\eqref{eq:z-scaling1}, reflecting the ambiguity of the dynamical scaling at
the Gaussian fixed point. The beta functions then become
\begin{align}
\frac{d a}{d \ln b} & = (2 - z - \eta_\psi) a + \frac{5}{2}
G\!\left(a^{-2}\right) g^2, \\
\frac{d g}{d \ln b} & = \frac{1}{2} (6 - d - z - \eta_\phi - 2 \eta_\psi) g
  + \frac{6}{5} H\!\left(a^{-2}\right) \frac{g^3}{a}, \label{eq:beta-g-2}\\
\frac{d \lambda}{d \ln b} & = \frac{1}{2} ( 6 - d - z - 3 \eta_\phi) \lambda
  - \frac{27}{4} \lambda^3 - \frac{\sqrt{3}}{35} \frac{g^3}{a},
\label{eq:beta-lambda-2}
\end{align}
with the anomalous dimensions
\begin{align}
 \eta_\psi & = \frac{4}{5} F\!\left(a^{-2}\right) \frac{g^2}{a}, \\
 \eta_\phi & = \frac{44}{35} \frac{g^2}{a} + \frac{21}{4} \lambda^2, \\
 \eta_2 & = \frac{2}{5} a g^2 + \frac{21}{4} \lambda^2,
\end{align}
and where we have rescaled $g$ and $\lambda$ as displayed below
Eq.~\eqref{eq:beta-lambda} in the main text and $\Lambda^{z + \eta_\psi - 2} a
\mapsto a$.
The functions $F$, $G$, and $H$ are also precisely the ones given in
Eqs.~\eqref{eq:F(c)}--\eqref{eq:H(c)} in the main text.

Starting the RG flow on the critical surface $r = 0$ in the vicinity of the
Gaussian fixed point, $\lambda \simeq 0$, $g \simeq 0$, we have initially $z
\simeq 1$, which renders the parameter $a$ relevant in the sense of the RG.
Together with $a$, however, $z$ increases towards the infrared and the flow of
$a$ will eventually stop when the dynamical exponent satisfies
\begin{equation} \label{eq:z-fixed-point2}
 z = 2 + \eta_\psi - \frac{5}{2} G\!\left(a^{-2}\right) \frac{g^2}{a}.
\end{equation}
Equating Eqs.~\eqref{eq:z-scaling2} and \eqref{eq:z-fixed-point2} determines
the value of $a$ at the interacting fixed point:
\begin{equation} \label{eq:fp-a}
 \left(2  - 2z - \frac{44}{35} \frac{g^2}{a}\right) \frac{1}{a^2} + \frac{2}{5}
\frac{g^2}{a} = 0.
\end{equation}
Upon rescaling $g^2/a \mapsto g^2$ and $\lambda^2 / a \mapsto \lambda$ the
Eq.~\eqref{eq:fp-a} becomes exactly the fixed-point equation for $c$
[Eq.~\eqref{eq:beta-c}] when $c = 1/a^2$.
With this identification, the flow equations for $g$ and $\lambda$ as well as
the anomalous dimensions and the dynamical exponent at the fixed point have
precisely the same form as in the main text, cf.\
Eqs.~\eqref{eq:beta-g}--\eqref{eq:z}
with Eqs.~\eqref{eq:beta-g-2}--\eqref{eq:z-fixed-point2}. The alternative
dynamical scaling scheme therefore leads to the same fixed-point
structure and universal critical exponents. The ambiguity in the dynamical
scaling is thus resolved at the QC point, which is determined by the unique
dynamical critical exponent
\begin{equation}
 z = 2 + \mathcal O(\epsilon^{6/5}).
\end{equation}

\section{Flow equations for four-dimensional tensor field}
\label{app:RG-four-dimensions}

We finally discuss the flow equations and fixed-point structure when evaluating
the angular integral directly at the upper critical dimension $d=4$ with the
nine $16\times 16$ gamma matrices $\gamma_a$, the $16$-component Dirac fermion
$\psi$, and the four-dimensional tensor field $T_{ij}$, $i,j=1,\dots,4$ with its
irreducible components $\phi_a$, $a=1,\dots,9$. The computation of the one-loop
diagrams in Fig.~\ref{fig:diagrams} now gives
\begin{align}
\frac{dc}{d\ln b} & = (2 - 2 z - \eta_{\phi} ) c
+ \frac{16}{9} g^2 + 9 \sqrt{c} \lambda^2,\\
\frac{d g}{d\ln b} & = \frac{1}{2} (6-d-z -\eta_{\phi} - 2 \eta_{\psi}) g +
\frac{28}{9} \tilde H(c) g^3, \\
\frac{d \lambda }{d\ln b} & = \frac{1}{2} (6-d-z - 3 \eta_{\phi}) \lambda
+ \frac{27}{2}\frac{\lambda^3}{\sqrt{c}} - \frac{1}{9}\sqrt{\frac{2}{3}} g^3,
\end{align}
with the anomalous dimensions
\begin{align}
\eta_{\psi} & = \frac{7}{6} F(c) g^2, \\
\eta_{\phi} & = \frac{49}{9} g^2 + 9\frac{\lambda^2}{\sqrt{c}}, \\
z & = 2 + \frac{9}{2} G(c) g^2 - \eta_{\psi}.
\end{align}
$F(c)$ and $G(c)$ are given in Eqs.~\eqref{eq:F(c)}--\eqref{eq:G(c)} in the main text and
\begin{equation}
\tilde H(c) = \frac{8+7\sqrt{c}}{8\left(1+\sqrt{c}\right)^2}.
\end{equation}
Out of the critical surface, the flow of the tuning parameter is
\begin{equation}
 \frac{d r}{d\ln b} = (2- \eta_\phi) r - \frac{64}{9} g^2 - 36
\frac{\lambda^2}{\sqrt{c}} \frac{1}{\left(1+r\right)^{3/2}}.
\end{equation}

The only qualitative and universal difference to the computation in $d=3$ is
the sign of the $\lambda^3$-term in $d \lambda / d \ln b$, which eliminates the
(unstable) purely bosonic fixed point (B in Fig.~\ref{fig:rg-flow}) at $g = 0$.
This is in full analogy to the Ginzburg-Landau-Wilson theory for the classical
nematic phase transition in liquid crystals, which exhibits a fixed point if and
only if the dimension $p$ of the tensor order parameter is $p < p_\mathrm c$
with $p_\mathrm c = 4$ to leading order in the related $\epsilon = 6-d$
expansion~\cite{lubenskypriest}. However, the existence of the fermionic fixed
point (F in Fig.~\ref{fig:rg-flow}) remains unaffected by this, and we find the
nontrivial solution for $c = \mathcal O(\epsilon^{6/5})$:
\begin{align}
\frac{\lambda_\pm^*}{c^{*1/4}} & = \pm \frac{1}{3}\sqrt{\epsilon}, &
\frac{g_\pm^*}{c^{*1/12}} & = \pm \sqrt{\frac{3}{2}}\sqrt{\epsilon},
\end{align}
where albeit $\lambda_\pm^*$ and $g_\pm^*$ now have the same sign. Examination
of the stability matrix shows that the fermionic fixed point now exhibits a
second relevant direction in direction of $\lambda$. This again reflects the
fact that
for the four-dimensional tensor order parameter there is no purely bosonic fixed
point at $g=0$ and $\lambda \neq 0$ and the flow in the direction of $\lambda$
is unbounded. In agreement with the discussion of the classical nematic phase
transition~\cite{lubenskypriest} we thus believe that the physical situation in
$d=3$ is more accurately described by the calculation directly in $d=3$ as
presented in the main text, which gives the stable fermionic fixed point with
$g^*$ and $\lambda^*$ being of opposite sign. In any case,
to the leading order we find for the $d=4$
calculation precisely the same values for the critical
exponents at the fermionic fixed point as in the main text [cf.\
Eqs.~\eqref{eq:exponents-fixed-point} and \eqref{eq:1-nu}],
\begin{align}
 \eta_\psi & = \mathcal O(\epsilon^{6/5}), &
 \eta_\phi & = \epsilon + \mathcal O(\epsilon^{6/5}), &
  z & = 2 + \mathcal O(\epsilon^{6/5}),
\end{align}
and
\begin{equation}
 1/\nu = 2 + 5 \epsilon + \mathcal O(\epsilon^{6/5}).
\end{equation}


\begin{thebibliography}{99}
\bibitem{note1} For the uniaxial states we can in fact evaluate the integral in
$E_\text{MF}$ [Eq.~\eqref{eq:mf-energy-1b}] analytically for any $\phi$: For
$\xi=0$, for instance, we find
$E_\text{MF}(\Lambda^2 \phi/g,0)/\Lambda^5 =
\frac{r}{2 \Lambda g^2} \phi^2
- \frac{4\pi}{(2\pi)^3} \Bigl[
- \frac{1}{6}
+ \frac{1}{4}\phi
+ \frac{1}{3}\phi^2
- \frac{\pi}{8} \phi^{5/2}
+ \frac{2 - 3 \phi + 6 \phi^2}{12\sqrt{3 \phi}} \artanh \frac{\sqrt{3\phi}}{1+\phi}
+ \frac{1}{8} \phi^{5/2} \arctan \sqrt{\frac{\phi}{(1-\phi)^2}}
+ \frac{5}{48\sqrt{3}} \phi^{5/2}
  \ln \frac{(1-\sqrt{3\phi}+\phi)^2}{1-\phi+\phi^2} \Bigr]$,
showing that $E_\text{MF}(\phi)$ at larger $\phi$
does not exhibit a second mininum and that the mean-field transition is indeed
continuous, as stated in the text.
\end{thebibliography}
\end{document}